\documentclass[preprint,showpacs,preprintnumbers,amsmath,amssymb]{revtex4}

\usepackage{graphicx}
\usepackage{dcolumn}
\usepackage{bm}

\begin{document}

\preprint{}
%
\begin{center}
{\bf Relativistic predictions of polarization phenomena in exclusive proton-induced proton-knockout reactions}\\
G.~C. Hillhouse$^{1}$, T. Ishida$^{2,3}$, T. Noro$^{2}$ and B.~I.~S. van der Ventel$^{1}$\\
$^{1}$Department of Physics, University of Stellenbosch, Stellenbosch, South Africa\\
$^{2}$Department of Physics, Kyushu University, Fukuoka, Japan
\end{center}
\footnotetext[3]{Present address: Laboratory of Nuclear Science, Tohoku University, Sendai 982-0826, Japan.}

\centerline{\bf ABSTRACT} 
Whereas a nonrelativistic distorted wave model fails to quantitatively describe analyzing power data 
for exclusive proton-induced proton-knockout from the 3$s_{1/2}$ state in $^{208}$Pb at 202~MeV, 
the corresponding relativistic prediction provides a perfect description, thus suggesting that the
Dirac equation is the more appropriate underlying dynamical equation. We check the consistency of this
result by comparing predictions for both dynamical models to new high resolution data for 
3$s_{1/2}$ knockout in $^{208}$Pb at a higher incident energy of 392~MeV.



\section{\label{sec:intro}Introduction}
\begin{figure}[htb]
\hspace{-1cm}\includegraphics[width=6cm,angle=90]{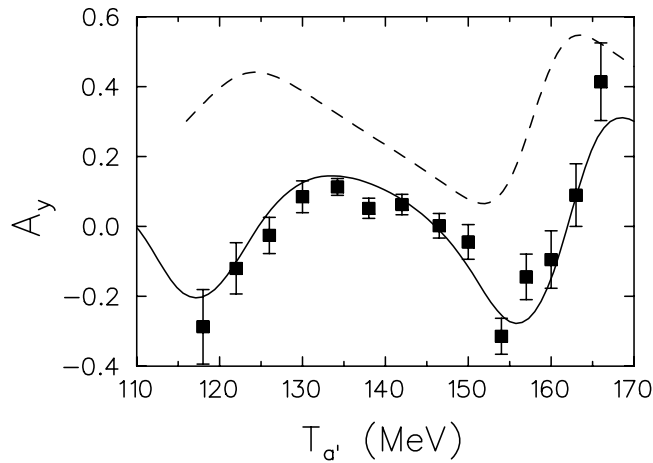}
\caption{\label{fig:ay-202MeV} 
Energy-sharing analyzing power A$_{y}$ for the knockout of protons from the 3s$_{1/2}$ state
in $^{208}$Pb, at an incident energy of 202~MeV, for coincident coplanar scattering angles 
($\theta_{a'} = 28.0^{\circ}$, $\theta_{b} = -54.6^{\circ}$), plotted as a function of the kinetic
energy of the proton scattered at angle $\theta_{a'}$. The dashed and solid curves represent the 
nonrelativistic and relativistic distorted wave predictions respectively. The data are from Ref. \cite{Ne02}.}
\end{figure}

Recently we demonstrated that a relativistic distorted wave model, represented by the solid 
line in Fig.~(\ref{fig:ay-202MeV}), provides a perfect description of the energy-sharing 
analyzing power data for exclusive proton knockout from the 3$s_{1/2}$ state in $^{208}$Pb 
at an incident energy of 202~MeV. The corresponding nonrelativistic prediction, indicated 
by the dashed line in Fig.~(\ref{fig:ay-202MeV}), completely fails to describe the data 
despite exhaustive corrections including different kinematic prescriptions for the NN amplitudes, 
non-local corrections to the scattering wave functions, and density-dependent modifications to 
the free NN scattering amplitudes \cite{Ne02,Hi03}. On the other hand, both Dirac and 
Schr\"{o}dinger-equation-based models provide an excellent description of the unpolarized 
energy-sharing cross section, thus highlighting the application of spin observables, such as 
the analyzing power, for discriminating between different dynamical effects in nuclear systems. 

\begin{figure}[htb]
\includegraphics[width=10cm]{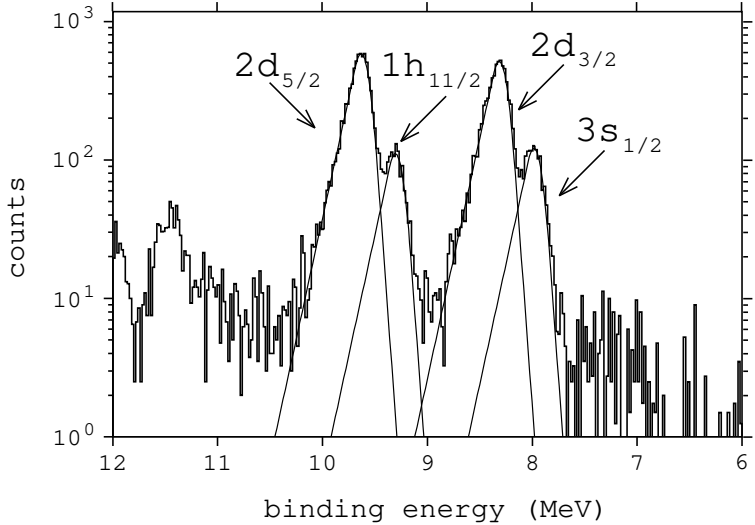}
\caption{\label{fig:resolution-392MeV} 
A typical binding-energy spectrum for the $^{208}$Pb($\vec{p}, 2p$)$^{207}$T$\ell$ reaction at 392~MeV.
The relevant single-particle proton-hole states in $^{208}$Pb are indicated. The data are from Ref.~\cite{Is06}.}
\end{figure}

Before claiming with absolute certainty that the relativistic Dirac equation is indeed the appropriate
underlying dynamical equation, it is essential to check the consistency of the 202~MeV result by 
considering 3$s_{1/2}$ knockout at another incident energy. In particular, one needs to evaluate 
to which extent the relatively poor energy resolution (310~KeV at FWHM), and resulting offline peak 
fitting analysis associated with the 202~MeV data, influences the quality of 3$s_{1/2}$ analyzing 
power data and interpretation thereof \cite{Ne02}. Indeed, the Kyushu University experimental nuclear 
physics group have recently performed a high resolution (250~KeV at FWHM) study of the $^{208}$Pb($\vec{p}, 2p$)$^{207}$T$\ell$ 
reaction at an incident energy of 392~MeV at the Research Center for Nuclear Physics in Japan \cite {Is06}: 
cross section and analyzing power data (energy-sharing and angular distributions) were measured for 
proton knockout from the 3$s_{1/2}$-, 2$d_{3/2}$, 2$d_{5/2}$ and 1$h_{11/2}$-states. A typical 
binding-energy spectrum is displayed in Fig.~(\ref{fig:resolution-392MeV}). 

In this paper, we focus on the theoretical interpretation of the 3$s_{1/2}$ cross section and analyzing power
data at 392~MeV. In particular we compare both relativistic and nonrelativistic distorted wave predictions to 
these new data so as to confirm or refute the previous claim at 202~MeV that relativistic dynamics are 
essential for a correct description of the 3$s_{1/2}$ analyzing power. Furthermore, we also identify additional 
independent polarization transfer observables which need to be measured to further study the role 
of relativity in nuclear reactions. Note that the original motivation for choosing a heavy 
target nucleus such as $^{208}$Pb was to maximize the influence of the nuclear medium of the scattering wave 
functions, while still maintaining the validity of the impulse approximation, and also avoiding complications 
associated with the inclusion of recoil corrections in the relativistic Dirac equation. In addition, for the 
knockout of 3$s_{1/2}$ valence protons one expects density-dependent corrections to the NN interaction to 
be negligible.

This paper is organized as follows: we briefly review the main aspects of the relativistic and nonrelativistic 
distorted wave models in Secs.~(\ref{sec:rdwia}) and (\ref{sec:nrdwia}) respectively. The relevant scattering 
observables are defined in Sec.~(\ref{sec:dsg-ay}), and results and conclusions are presented in 
Sec.~(\ref{sec:results-and-conclusions}).

\section{\label{sec:rdwia}Relativistic model}

The formalism for the relativistic distorted wave model has been presented in 
Refs.~\cite{Ik95,Ma96,Ma98,Hi03,Hi03b}. In this paper we briefly allude to the 
most important aspects of the model and refer the interested reader to the latter 
references for more detail. 

The exclusive ($p,2p$) reaction of interest is schematically depicted in Fig.~(\ref{fig-p2pgeometry}), 
whereby an incident proton, $a$, knocks out a bound proton, $b$, from a specific orbital in the 
target nucleus $A$, resulting in three particles in the final state, namely the recoil residual 
nucleus, $C$, and two outgoing protons, $a'$ and $b$, which are detected in coincidence at coplanar 
laboratory scattering angles, $\theta_{a'}$ and $\theta_{b}$, respectively. All kinematic quantities 
are completely determined by specifying the rest masses, $m_{i}$, of particles, where $i$ = ($a$,$A$, 
$a'$, $b$, $C$), the laboratory kinetic energy, $T_{a}$, of incident particle $a$, the laboratory kinetic 
energy, $T_{a'}$, of scattered particle $a'$, the laboratory scattering angles $\theta_{a'}$ and $\theta_{b}$, 
and also the excitation energy of the proton that is to be knocked out of the target nucleus, $A$.

\begin{figure}
\includegraphics[width=12cm]{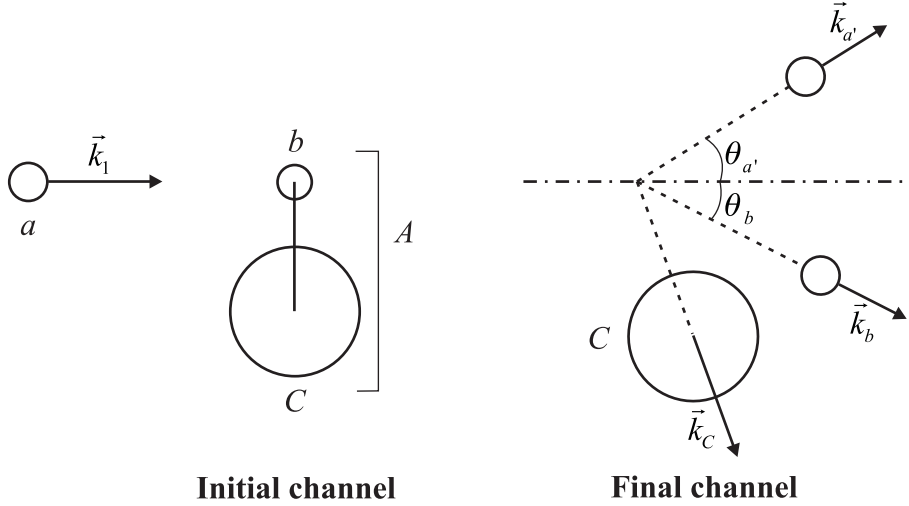}
\caption{\label{fig-p2pgeometry}Schematic representation for the 
coplanar $(p,2p)$ reaction of interest.}
\label{fig-p2pgeometry}
\end{figure}

We adopt a zero-range approximation to the NN interaction, whereby the relativistic distorted wave transition 
matrix element is given by
\begin{eqnarray}
\hspace{-7mm} T_{L J M_J}(s_{a}, s_{a'}, s_{b}) & = & 
\int d\vec{r}\, 
[\,
\bar{\Psi}^{(-)}(\vec{r}, \vec{k}_{a'C},s_{a'}) 
\otimes   
\bar{\Psi}^{(-)}(\vec{r}, \vec{k}_{bC},s_{b})
\,] \nonumber\\
 & & \hat{F}_{NN}(T_{\rm eff}^{\rm \ell ab}, \theta_{\rm eff}^{\rm cm})\,
[\,
\Psi^{(+)}(\vec{r}, \vec{k}_{aA}, s_{a})
\otimes
\Phi^{B}_{L J M_J}(\vec{r}\,)
\,]
\label{e-tjlm}
\end{eqnarray}
where $\otimes$ denotes a Kronecker product. The four-component scattering wave functions, 
$\psi(\vec{r}, \vec{k}_{i},s_{i})$, are solutions to the fixed-energy Dirac equation with 
spherical scalar and time-like vector nuclear optical potentials: $\Psi^{(+)}(\vec{r}, \vec{k}_{aA},s_{a})$ 
is the relativistic scattering  wave function of the incident particle, $a$, with outgoing boundary 
conditions [indicated by the superscript $(+)$], where $\vec{k}_{aA}$ is the momentum of particle 
$a$ in the ($a$ + $A$) center-of-mass system, and $s_{a}$ is the spin projection of particle $a$ 
with respect to $\vec{k}_{aA}$ as the $\hat{z}$-quantization axis;  $\bar{\Psi}^{(-)}(\vec{r}, \vec{k}_{j C},s_{j})$
 is the adjoint relativistic scattering wave function for particle $j$ [ $j$ = ($a',b$)] with incoming 
boundary conditions [indicated by the superscript $(-)$], where $\vec{k}_{j C}$ is the momentum of particle
$j$ in the ($j$ + $C$) center-of-mass system, and $s_{j}$ is the spin projection of particle $j$ with 
respect to $\vec{k}_{j C}$ as the $\hat{z}$-quantization axis. We employ a global scalar and vector Dirac 
optical potential parameter set for the distorting optical potentials. More specifically, we use the 
energy-dependent mass-independent ``EDAI-fit'' parameter set which has been constrained by proton 
elastic scattering data on $^{208}$Pb for incident proton energies between 21 MeV and 1040~MeV \cite{Co93b}. 
The four-component relativistic boundstate proton wave function, which is denoted by 
$\Phi^{B}_{L J M_J}(\vec{r}\, )$ in Eq.~(\ref{e-tjlm}) and labeled 
by single-particle quantum numbers $L$, $J$, and $M_{J}$, is obtained via selfconsistent solution to the 
Dirac-Hartree field equations within the context of the relativistic mean field approximation associated 
with the QHDII Lagrangian density \cite{Ho81} of quantum hadrodynamics \cite{Ho91}. We also employ 
the impulse approximation which assumes that the form of 
the NN scattering matrix in the nuclear medium is the same as that for free NN scattering. In addition, we adopt 
the IA1 representation \cite{Mc83a} which parameterizes the NN scattering matrix $\hat{F}_{NN}(T_{\rm eff}^{\rm \ell ab}, 
\theta_{\rm eff}^{\rm cm})$ in terms of five Lorentz invariant amplitudes (scalar, pseudoscalar, vector, axial-vector, 
tensor) which are directly related to the five nonrelativistic Wolfenstein amplitudes describing on-shell 
NN experimental scattering data: $T_{\rm eff}^{\rm \ell ab}$ and $\theta_{\rm eff}^{\rm cm}$ represent the effective 
two-body laboratory kinetic energy and center-of-mass scattering angles, respectively, based 
on the so-called final-energy prescription \cite{Ch83}. 

\section{\label{sec:nrdwia}Nonrelativistic model}
All of the nonrelativistic calculations are based on the computer code {\sc threedee} of Chant and Roos \cite{Ch83},
where the nonrelativistic transition amplitude, based on a zero-range approximation to the NN interaction, is
\begin{eqnarray}
\hspace{-7mm} T_{L J M_J}^{\mbox{\scriptsize NRDW}}
(s_{a}, s_{a'}, s_{b})  &=&  \int d\vec{r}\, 
[\,
\psi^{*(-)}(\vec{r}, \vec{k}_{a'C},s_{a'}) 
\otimes   
\psi^{*(-)}(\vec{r}, \vec{k}_{bC},s_{b})
\,] \nonumber\\
& &  
\hat{t}_{NN}(T_{\rm eff}^{\rm \ell ab}, 
\theta_{\rm eff}^{\rm cm})
[\,
\psi^{(+)}(\gamma\vec{r}, \vec{k}_{aA}, s_{a})
\otimes
\varphi^{B}_{L J M_J}(\vec{r}\,)
\,]
\label{nr-tjlm},
\end{eqnarray}
where the $\psi$'s represent the appropriate incoming and outgoing nonrelativistic two-component 
scattering wave functions and $\varphi$ is the nonrelativistic wave function of the bound proton to 
be knocked out, and $\gamma=A/(A+1)$, where $A$ being the target mass number: both scattering and 
boundstate wave functions are solutions to the Schr\"{o}dinger equation. For consistency between 
the relativistic and non-relativistic calculations, the nonrelativistic radial boundstate wave function 
is approximated by taking the upper component radial wave function of the relativistic four component 
boundstate wave function employed in the relativistic predictions. The scattering wave functions are 
solutions to the nonrelativistic Schr\"{o}dinger equation employinga Schr\"{o}dinger-equivalent 
representation \cite{Co93b} of the relativistic global optical potentials mentioned in Sec.~(\ref{sec:rdwia}). I
n this paper we employ experimental amplitudes for the NN scattering matrix $\hat{t}_{NN}$ which are 
determined from the NN Arndt phase shift analysis (January 1999) \cite{Ar00}: these amplitudes are directly 
related to the relativistic Lorentz invariant amplitudes \cite{Mc83a}.

\section{\label{sec:dsg-ay}Scattering observables}
The spin observables of interest are denoted by $\mbox{D}_{i' j}$ and are 
related to the probability that an incident beam of particles $a$ 
with spin-polarization $j$ induces a spin-polarization $i'$ for 
the scattered beam of particles $a'$: the subscript $j = (0,\ell,n,s)$ 
is used to specify the polarization of the incident beam, $a$, along any 
of the orthogonal directions
\begin{eqnarray}
\hat{\ell}\ =\ \hat{z}\ =\ \hat{k}_{aA}\, ,\ \ \ \hat{n}\ =\ \hat{y}\ =\ \hat{k}_{aA} \times \hat{k}_{a' C}\, , \ \ \ 
\hat{s}\ =\ \hat{x}\ =\ \hat{n} \times \hat{\ell}\,,
\label{e-lns}
\end{eqnarray}
and the subscript $i' = (0,\ell',n',s')$ denotes the polarization of the scattered 
beam, $a'$, along any of the orthogonal directions:
\begin{eqnarray}
\hat{\ell}'\ =\ \hat{z}'\ =\ \hat{k}_{a' C}\, , \ \ \ \hat{n}'\ =\  \hat{n}\ =\ \hat{y}\, ,\ \ \
\hat{s}'\ =\ \hat{x}'\ =\ \hat{n} \times \hat{\ell}'\,.
\label{e-lpnsp}
\end{eqnarray}
The choice $j\,(i') = 0$ is used to denote an unpolarized incident (scattered) beam.
With the above coordinate axes in the initial and final channels, the 
spin observables, $\mbox{D}_{i' j}$, are defined by
\begin{eqnarray}
\mbox{D}_{i' j}\ =\ \frac{ \sum_{M_J, s_b}\, \mbox{Tr}(T \sigma_{j} T^{\dagger} \sigma_{i'})}
{ \sum_{M_J, s_b}\, \mbox{Tr}(T T^{\dagger})}\,,
\label{e-dipj}
\end{eqnarray}
where $\mbox{D}_{n 0}=\mbox{P}$ refers to the induced polarization, 
$\mbox{D}_{0 n}=\mbox{A}_{y}$ denotes the analyzing power, and the other
polarization transfer observables of interest are
$\mbox{D}_{n n},\, \mbox{D}_{s' s},\, \mbox{D}_{\ell' \ell},\, 
\mbox{D}_{s' \ell},\ \mbox{and}\ \mbox{D}_{\ell' s}$. 
The denominator of Eq.~(\ref{e-dipj}) is related to the unpolarized triple 
differential cross section, i.e.,
\begin{eqnarray}
\sigma\ \ =\ \
\frac{d^3 \sigma}
{d T_{a'}\, d \Omega_{a'}\, d \Omega_b}\ \ =\ \  S_{L J}\, \sigma_{\rm{calc}}\, ,\\
\sigma_{\rm{calc}}\ \ =\ \ \frac{F_{\rm{kin}}}{(2s_{a} + 1)\, (2 J + 1)} \sum_{M_J, s_b}\, \mbox{Tr}(T\, T^{\dagger})\,
\label{e-unpol}
\end{eqnarray}
where $F_{\rm{kin}}$ is a kinematic factor and $S_{L J}$ is the spectroscopic factor \cite{Ma96,Ma98}.
In Eq.~(\ref{e-dipj}), the symbols $\sigma_{i'}$ and $\sigma_{j}$ denote the usual $2 \times 2$ Pauli 
spin matrices, and the $2 \times 2$ matrix $T$ is given by
\begin{eqnarray}
T\ =\ \left(
\begin{array}{cc}
T_{L J}^{s_{a} = +\frac{1}{2}, s_{a'} = +\frac{1}{2}} & 
T_{L J}^{s_{a} = -\frac{1}{2}, s_{a'} = +\frac{1}{2}} \\
T_{L J}^{s_{a} = +\frac{1}{2}, s_{a'} = -\frac{1}{2}} & 
T_{L J}^{s_{a} = -\frac{1}{2}, s_{a'} = -\frac{1}{2}} 
\end{array}\right)
\label{e-ttwobytwo}
\end{eqnarray}
where $s_{a} = \pm \frac{1}{2}$ and $s_{a'} = \pm \frac{1}{2}$ refer to the spin projections of particles 
$a$ and $a'$ along the $\hat{z}$ and $\hat{z}'$ axes, defined in Eqs.~(\ref{e-lns}) and (\ref{e-lpnsp}), 
respectively; the matrix $T_{L J}^{s_{a}, s_{a'}}$ is related to the relativistic $(p,2p)$ transition
matrix element $T_{L J M_{J}}(s_{a}, s_{a'}, s_{b})$, defined in Eq.~(\ref{e-tjlm}) 
via
\begin{eqnarray}
T_{L J}^{s_{a}, s_{a'}}\ =\ T_{L J M_J}
(s_a, s_{a'}, s_b)\,.
\label{e-tljmrelatet}
\end{eqnarray}

\section{\label{sec:results-and-conclusions}Results and conclusions}
The experimental data (experiment E205) were measured using the dual-arm spectrometer at RCNP \cite{Is06}. 
For a direct comparison to data both relativistic (RDWIA)and nonrelativistic (NRDWIA) distorted wave predictions 
were corrected (using a Monte-Carlo simulation) for the finite angle acceptance of both Grand Raiden and the Large 
Acceptance Spectrometers: note that the current version of the relativistic code does not consider out-of-plane 
predictions, and consequently the RDWIA calculations exclude finite azimuthal angle $\phi$ corrections to the solid angle. 

\begin{figure}[htb]
\includegraphics[width=9.5cm,angle=90]{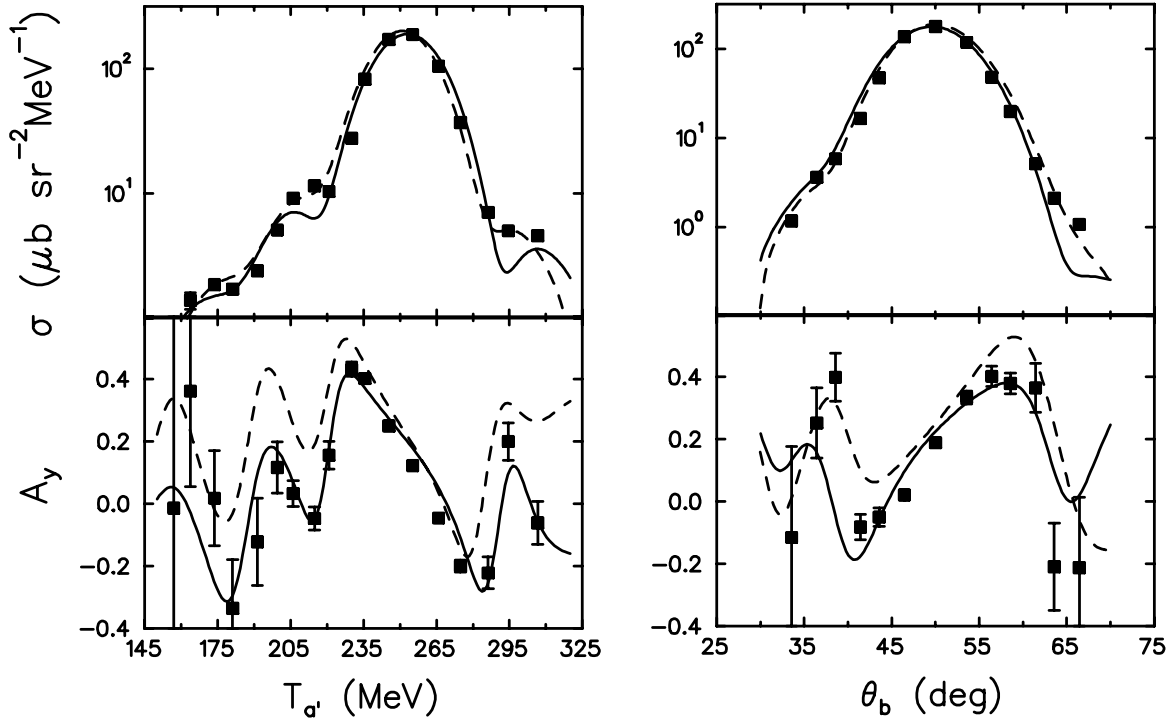}
\caption{\label{fig:ay-dsg-392MeV} 
Unpolarized triple differential cross section $\sigma$ and analyzing power A$_{y}$ for proton knockout from the 
3$s_{1/2}$ state in $^{208}$Pb at an incident energy of 392~MeV plotted as a function of the 
kinetic energy $T_{a'}$ for coincident coplanar laboratory scattering angles ($32.5^{\circ}$, $-50.0^{\circ}$) 
(left panel), and as a function of the scattering angle $\theta_{b}$, for $T_{a'} = 250$~MeV and 
$\theta_{a'} =  32.5^{\circ}$ (right panel). The dashed and solid curves represent the nonrelativistic 
and relativistic distorted wave predictions respectively. The data are from Ref. \cite{Is06}.}
\end{figure}

We now compare theoretical RDWIA and NRDWIA energy-sharing and angular distribution predictions of cross sections 
$\sigma$ and analyzing powers A$_{y}$ to experimental data for 3$s_{1/2}$ knockout from $^{208}$Pb at 392~MeV. 
In Fig.~(\ref{fig:ay-dsg-392MeV}) the energy-sharing distributions (left panel) are plotted as a function of the 
kinetic energy $T_{a'}$ for coincident coplanar laboratory scattering angles ($32.5^{\circ}$, $-50.0^{\circ}$), and the 
angular distributions (right panel) are plotted as a function of the scattering angle $\theta_{b}$, for $T_{a'} = 250$~MeV 
and $\theta_{a'} =  32.5^{\circ}$. Both RDWIA (solid curves) and NRDWIA (dashed curves) models provide a satisfactory 
description of the shape of the unpolarized cross sections for both energy-sharing and angular distributions. The analysis 
regarding the extraction of spectroscopic factors $S_{LJ}$ and corresponding error bars is currently in progress: the 
values of $S_{LJ}$ represent single-particle state occupation numbers and are obtained by normalizing the 
calculated cross sections $\sigma_{\rm{calc}}$ [see Eq.~(\ref{e-unpol})] to the experimental cross section data.

Next we turn our attention to the analyzing power. As is the case for the 202~MeV data, the RDWIA energy-sharing 
distribution, indicated by the solid line in Fig.~(\ref{fig:ay-dsg-392MeV}), is significantly reduced compared 
to the corresponding NRDWIA calculations (dashed line). Furthermore, the RDWIA prediction provides a better overall 
quantitative description of the data. Recall that for the 202~MeV data the energy-sharing analyzing power distribution 
for RDWIA was consistently reduced compared to the RDWIA calculations for all values of $T_{a'}$. The latter is also
true at 392~MeV, except at the maximum value of the cross section (at $T_{a'} \approx 245$~MeV) where both RDWIA and 
NRDWIA describe the data equally well. The RDWIA angular distribution is also consistently reduced compared to the 
NRDWIA result, except at $\theta_{b} \approx 50^{\circ}$ (corresponding to the maximum value of the cross section)
where both models descrribe the data equally well. Hence, in general we conclude that at both 202 and 392~MeV, the 
RDWIA model is superior compared to the NRDWIA model, thus suggesting that relativistic dynamics are important for 
for describing $(p,2p$) reactions. 

\begin{figure}[htb]
\includegraphics[width=11cm,angle=90]{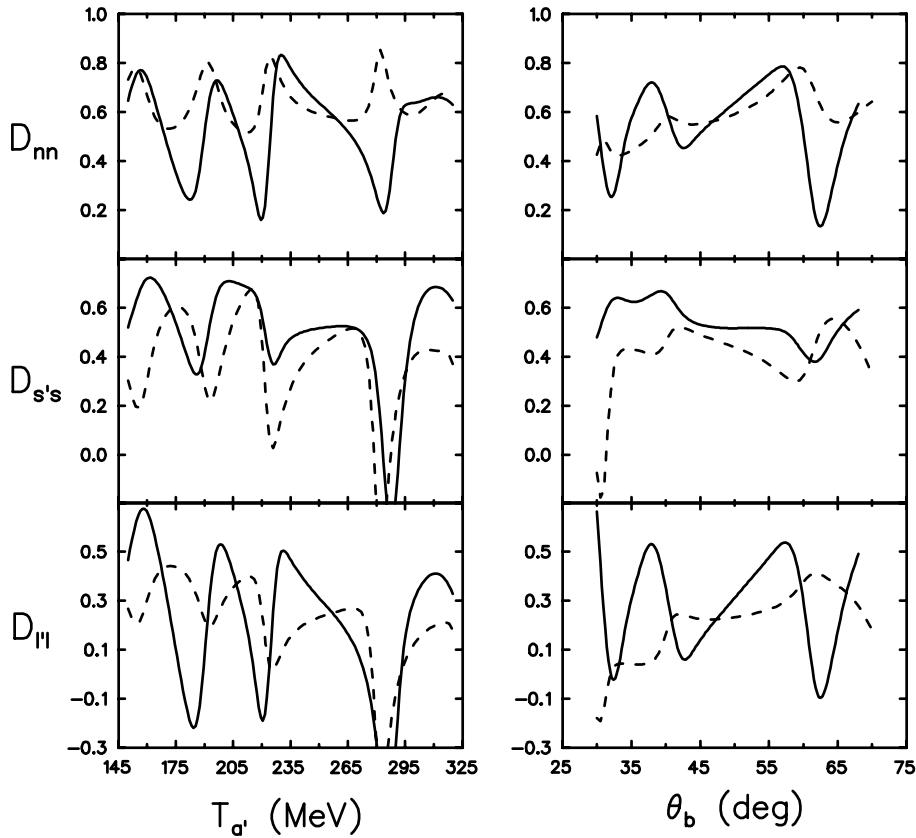}
\caption{\label{fig:dij-rel-vs-nr-392MeV} 
Polarization transfer observables, D$_{nn}$, D$_{s's}$ and D$_{\ell ' \ell}$, plotted as a function of 
either $T_{a'}$ (left panel) or $\theta_{b}$ (right panel) for kinematics corresponding to observables
plotted in the left and right panels of Fig.~(\ref{fig:ay-dsg-392MeV}) respectively. The dashed and solid 
curves represent the nonrelativistic and relativistic distorted wave predictions respectively.}
\end{figure}

It is desirable to check the consistency of this result by also measuring other polarization observables which are 
sensitive to differences between relativistic versus nonrelativistic dynamical models. Based on the
same kinematic conditions as for the 392~MeV analyzing power, we also identify the spin observables, D$_{nn}$, 
D$_{s' s}$ and D$_{\ell ' \ell}$ as good candidates for further studying the role of relativity in nuclei: 
see Fig.~(\ref{fig:dij-rel-vs-nr-392MeV}): Other spin observables are not displayed since they are less
sensitive to different dynamical effect.

The success of the RDWIA model to describe the knockout of the 3$s_{1/2}$ valence nucleons in $^{208}$Pb
inspires confidence to extend our model to systematically address the topical question of how the free NN
interaction is modified by the presence of the neighbouring nucleons in nuclei: the exclusive nature of 
($p,2p$) reactions allows one to selectively knockout protons from deep- to low-lying single-particle 
orbitals in nuclei, thus enabling one to systematically study the effect of the nuclear density on 
the NN interaction.  In addition, we intend to apply our RDWIA model to study exclusive proton knockout 
reactions from proton-rich and neutron-rich exotic nuclei using inverse kinematics at future radioactive 
beam facilities such as RIKEN and GSI. A drawback of the current implementation of the RDWIA model is 
the use of the ambiguous IA1 parameterization for the NN scattering matrix. Future work will include 
study the effect of replacing the IA1 with the unambiguous IA2 representation in terms of 44 independent 
invariant amplitudes (of which the IA1 representation is a subset) which are consistent with parity and 
time-reversal invariance as well as charge\\
symmetry \cite{Tj87}.

\begin{acknowledgements}
We acknowledge support from the South African National Research Foundation under 
grant numbers 2054166 (GCH) and 2948567 (BISvdV). GCH also acknowledges support from 
the Japan Society for the Promotion of Science (GCH).
\end{acknowledgements}

\end{document}